\documentclass[11pt]{article}
\usepackage{graphicx}

\title{On the Shoulders of Giants: The Growing Impact of Older Articles}

\author{
Alex Verstak, Anurag Acharya, Helder Suzuki, Sean Henderson,\\
Mikhail Iakhiaev, Cliff Chiung Yu Lin,  Namit Shetty\\
\\
Google Inc.
}

\begin{document}

\maketitle

\begin{abstract}

In this paper, we examine the evolution of the impact of older
scholarly articles.  We attempt to answer four questions. First, how
often are older articles cited in scholarly papers and how has this
changed over time. Second, how does the impact of older articles vary
across different fields of scholarship. Third, is the change in the
impact of older articles accelerating or slowing down. Fourth, are
these trends different for much older articles. 

To answer these questions, we studied citations from articles
published in 1990-2013. We computed the fraction of citations to older
articles from articles published each year as the measure of impact
for the study. For this study, we considered articles that were
published at least 10 years before the citing article as {\it older
  articles}. To explore how changes in citation behavior differ across
areas of research, we computed these numbers for 261 subject
categories and 9 broad areas of research. Finally, we repeated the
computation for two other definitions of older articles, 15 years and
older and 20 years and older.

There are three major conclusions from our study. First, the impact of
older articles has grown substantially over 1990-2013. Our analysis
indicates that, in 2013, 36\% of citations were to articles that are
at least 10 years old and that this fraction has grown 28\% since
1990. The fraction of older citations increased over 1990-2013 for 7
out of 9 broad areas of research and 231 out of 261 subject
categories.

Second, the change over the second half (2002-2013) was significantly
larger than that over first half (1990-2001) --- the increase in the
second half was double the increase in the first half.

Third, the trend of a growing impact of older articles also holds for
articles that are at least 15 years old and those that are at least 20
years old. In 2013, 21\% of citations were to articles $\geq 15$ years
old with an increase of 30\% since 1990 and 13\% of citations were to
articles $\geq 20$ years old with an increase of 36\% over the same
period.

Now that finding and reading relevant older articles is about as easy
as finding and reading recently published articles, significant
advances aren't getting lost on the shelves and are influencing work
worldwide for years after.

\end{abstract}

\section{Introduction}

The last two decades have seen several dramatic changes in scholarly
communication. First, scholarly journals have largely moved from
physical distribution of print issues to online availability of
individual articles. A large number of journals have also digitized
older articles and made them available online. Second, search services
now index the entire text of articles instead of just abstracts and
keywords. The common ranking approach has moved from reverse
chronological (most-recent-first) to relevance ranking
(most-relevant-first). Third, many journals now make articles
available much sooner - often, soon after acceptance. Furthermore,
several disciplines have developed large preprint collections which
include articles before they have been accepted for formal
publication. Consequently, researchers can learn about new results
much sooner than was possible earlier. Fourth, the number of articles
and journals has grown rapidly -- between 1990 and 2013, the number of
scholarly articles published per year grew close to 3-fold. As a
result, there is much more recent work for researchers to learn from,
build upon and cite.

The first two changes have made it easier for researchers to find the
most relevant articles for their work regardless of the age of the
articles. Finding and reading relevant older articles is now about as
easy as finding and reading recently published articles. If these were
the only changes, it would be reasonable to expect that the fraction
of citations to older articles would increase. 

The second set of changes have greatly expanded the amount of
concurrent and recent work that researchers need to situate their work
in relation to. If these were the only changes, it would be reasonable
to expect that fraction of citations to recent articles would increase
and the fraction of citations to older articles would decrease.

To understand the evolution of the impact of older scholarly articles,
we studied citations from articles that were published in
1990-2013. We attempted to answer four questions. First, how often are
older articles cited in scholarly papers and how has this changed over
time. Second, how does the impact of older articles vary across
different fields of scholarship. Third, is the change in the impact of
older articles accelerating or slowing down. Fourth, are these trends
different for much older articles.

We computed the fraction of citations to older articles from articles
published each year as the measure of impact for the study. For this
study, we considered articles that were published at least 10 years
before the citing article as {\it older articles}.  To explore how
changes in citation behavior differ across areas of research, we
computed these numbers for 261 subject categories and 9 broad areas of
research. Finally, we repeated the computation for two other
definitions of older articles, 15 years and older and 20 years and
older.

There are three major conclusions from our study. First, the impact of
older articles, as measured by citations, has grown substantially over
1990-2013. Our analysis indicates that, in 2013, 36\% of citations
were to articles that are at least 10 years old and that this fraction
has grown 28\% since 1990. While there were field-specific variation
in the evolution, the fraction of older citations increased over
1990-2013 for 7 out of 9 broad areas of research and 231 out of 261
subject categories.

Second, the change over the second half (2002-2013) was significantly
larger than that over first half (1990-2001) --- the increase in the
second half was double the increase in the first half. For context,
most archival digitization efforts as well as the move to fulltext
relevance-ranked search occurred over the second half (2002-2013).

Third, the trend of a growing impact of older articles also holds for
articles that are at least 15 years old and those that are at least 20
years old. In 2013, 21\% of citations were to articles $\geq 15$ years
old with an increase of 30\% since 1990 and 13\% of citations were to
articles $\geq 20$ years old with an increase of 36\% over the same
period.

Now that finding and reading relevant older articles is about as easy
as finding and reading recently published articles, significant
advances aren't getting lost on the shelves and are influencing work
worldwide for years after.

For the rest of this paper, we refer to citations to older articles as
{\it older citations} for the sake of brevity. In the following
section, we describe the analysis steps. Next, we present and discuss
the results. After that, we describe related work.

\section{Methods}

For this study, we included all journals and conferences that were
assigned to one or more categories in the 2014 release of Scholar
Metrics. The Scholar Metrics inclusion criteria for publication venues
were~\cite{scholar-metrics2014}: (1) publish 100 or more articles over
2009-2013, (2) at least one article must receive at least one citation
over 2009-2013, (3) follow Google Scholar indexing guidelines. Scholar
Metrics limits categorization into subject categories to English
publications.  Accordingly, this study covers all the English language
journals and conferences included in Scholar Metrics. Scholar Metrics
displays up to 20 top publications per subject category, and makes the
remaining ones available via keyword search. This study covers {\it
  all} the categorized journals and conferences, not only the top 20
per category. Scholar Metrics also includes selected preprint
repositories. Preprint repositories are not included in this study.

We used all the 261 subject categories from the 2014 release of
Scholar Metrics. To explore trends in broad areas, we also grouped
subject categories into nine broad research areas. We used the broad
areas from Scholar Metrics for this, with one change --- we separated
{\it Engineering} and {\it Computer Science}. The citation patterns in
these two areas are significantly different and this separation
allowed us to explore the differences. We also added {\it All
  articles} as the union of all broad areas.

We created a group of articles for each subject-category-year and
broad-area-year combination, such as {\it Immunology} for the year
2000 or {\it Physics \& Mathematics} for the year 2004. Each
category-year/area-year group included all articles published in the
given year in all publications in the given category/area.

For each publication, we included all articles with a publication date
within 1990-2013, both inclusive. Note that each journal or conference
can be associated with more than one subject category. Such
publications are included in the computation for each category they
are a part of.

For each category-year/area-year group, we computed the total number
of citations as well as the number of citations to articles published
in each preceding year. These citation counts (total citations as well
as the number of citations for each preceding year) included {\it all}
the citations from these articles, not just the citations to articles
included in this study. We used this matrix to compute the fraction of
citations to older articles. We used three different thresholds for
older articles, $\geq 10$ years old, $\geq 15$ years old, and $\geq
20$ years old.

To see if the rate of change in the fraction of older citations is
speeding up or slowing down, we computed the aggregate change for
1990-2001 (first half) and 2002-2013 (second half) for every category.

\section{Results}

\begin{figure}
\centering
\includegraphics{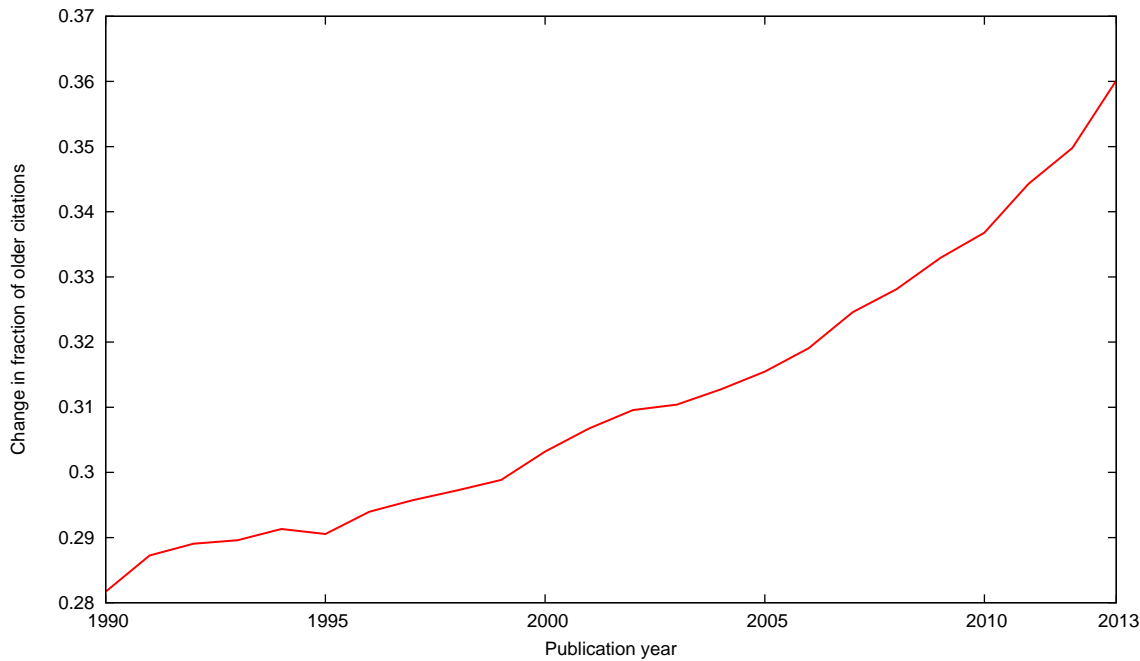} 
\caption{Fraction of older citations from all papers published in 1990-2013.}
\label{fig:all-10-year} 
\end{figure}

Figure~\ref{fig:all-10-year} presents the evolution of the fraction of
citations to articles that are at least ten years old. It covers all
publications included in this study. It shows that the fraction of
older citations has grown steadily over 1990-2013. It also shows that
the growth rate was roughly fixed over 1990-1999 and has accelerated
after that.

\begin{figure}
\centering
\includegraphics{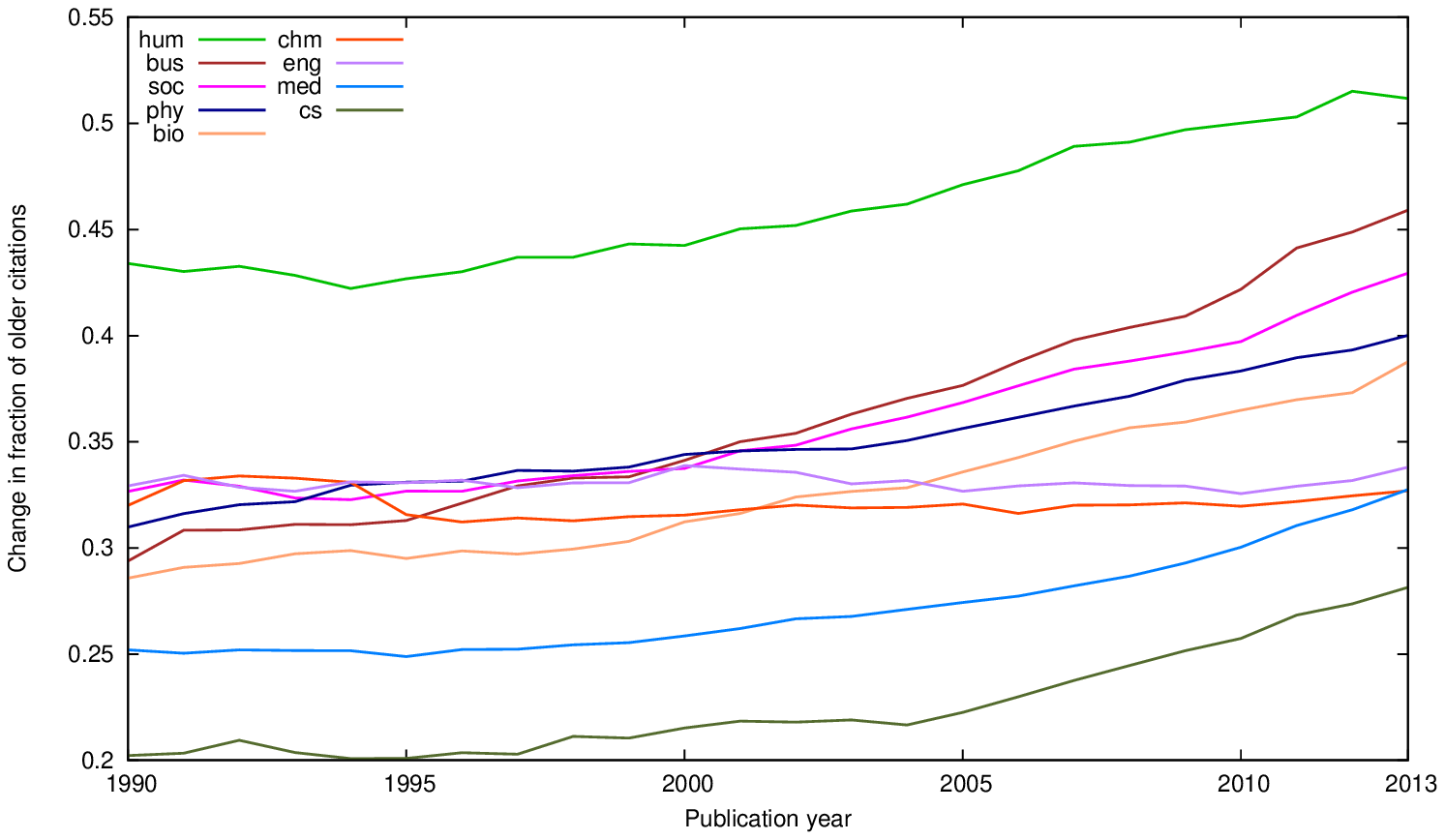} \\
{\small
{\bf bio:} Life Sciences \& Earth Sciences;
{\bf bus:} Business, Economics \& Management;
{\bf cs:} Computer Science;
{\bf chm:} Chemical \& Material Sciences;
{\bf eng:} Engineering;
{\bf hum:} Humanities, Literature \& Arts;
{\bf med:} Health \& Medical Sciences;
{\bf phy:} Physics \& Mathematics;
{\bf soc:} Social Sciences}\\
\caption{Fraction of citations to older articles for broad areas of research. }
\label{fig:broad-10-year} 
\end{figure}

Figure~\ref{fig:broad-10-year} presents the evolution of the fraction
of older citations for all broad areas. It shows that 7 out of 9
broad areas saw a substantial increase in the fraction of
older citations over 1990-2013. Two broad areas, {\it Chemical \&
  Material Sciences} and {\it Engineering}, did not see a significant
change in the fraction of older citations.

Table~\ref{tab:broad-frac-and-growth} presents the fraction of older
citations as well as the change since 1990 in numerical form (for ease
of comparison). The change over 1990-2013 is computed
as a percentage:\\

$(fraction\_in\_2013-fraction\_in\_1990)/fraction\_in\_1990 * 100$\\

It shows that, in 2013, four broad areas had at least 40\% citations
to older articles, {\it Humanities, Literature \& Arts} being the
highest at 51\%. Five broad areas had an increase of over 30\% in the
fraction of older citations over 1990-2013, the highest being the 56\%
growth seen by {\it Business, Economics \& Management}.

\begin{table}
\centering
\begin{tabular}{|l|r|r|r|}
\hline
Broad area & Older citations in 2013 & Change since 1990 \\  \hline
Humanities, Literature \& Arts & 51\% & 18\%\\
Business, Economics \& Management & 46\% & 56\%\\
Social Sciences & 43\% & 31\%\\
Physics \& Mathematics & 40\% & 29\%\\
Life Sciences \& Earth Sciences & 39\% & 36\%\\
Engineering & 34\% & 3\%\\
Chemical \& Material Sciences & 33\% & 2\%\\
Health \& Medical Sciences & 33\% & 30\%\\
Computer Science & 28\% & 39\%\\
{\it All articles} & {\it 36\%} & {\it 28\%}\\
\hline
\end{tabular}
\caption{Change in the fraction of older citations over 1990-2013.}
\label{tab:broad-frac-and-growth}
\end{table}

Table~\ref{tab:broad-change-histogram} presents a histogram of the
variation in the growth in the fraction of older citations for each of
the broad categories. Buckets in the histogram count the number of
subject categories whose growth is within a given range. It shows
that, overall, 231 out of 261 categories (89\%) saw an increase in the
fraction of older citations. That is, the growth in the fraction of
older articles has occurred over diverse fields - which vary greatly
in terms of their publication frequency and citation patterns.

Looking more closely, it shows that 102 out of 261 subject categories
saw a growth in the fraction of older citations that was over 30\%, 44
of them with an increase over 50\%. For two broad areas, {\it
  Business, Economics \& Management} and {\it Computer Science}, a
little less than two-thirds of the subject categories (10 out of 16
and 11 out of 18 respectively) saw a growth over 50\% in the fraction
of older citations. It also shows that all broad areas had some
subject categories with $> 50\%$ increase in the fraction of older
citations. Finally, it shows that a large fraction of the subject
categories that saw a drop in the fraction of older citations are a
part of {\it Chemical \& Material Sciences} and {\it Engineering}.

Note that some subject categories are included in more than one broad
category. For example, {\it Development Economics} and {\it Human
  Resources \& Organizations} are a part of both {\it Business,
  Economics \& Management} and {\it Social Sciences}. As a result, the
sum of the counts in a column in
Table~\ref{tab:broad-change-histogram}, e.g., number of subject
categories that saw 0-20\% growth, is expected to be larger than the
number in the same column for {\it All articles}.

\begin{table}
\centering
\begin{tabular}{|l|r|r|r|r|r|r|}\hline
Broad area & $< 0\%$ & 0-20\% & 20-30\% & 30-40\% & 40-50\% & $> 50\%$\\ \hline
Humanities, Literature \& Arts & 2 & 10 & 5 & 5 & 2 & 2\\
Business, Economics \& Management & 0 & 0 & 3 & 1 & 2 & 10 \\
Social Sciences & 4 & 10 & 11 & 8 & 9 & 10\\
Physics \& Mathematics & 2 & 3 & 10 & 4 & 2 & 3\\
Life Sciences \& Earth Sciences & 6 & 10 & 9 & 7 & 2 & 5\\
Engineering & 10 & 15 & 4 & 7 & 2 & 2\\ 
Chemical \& Material Sciences & 11 & 5 & 1 & 0 & 0 & 2\\
Health \& Medical Sciences & 3 & 30 & 17 & 8 & 2 & 9\\
Computer Science & 1 & 2 & 2 & 0 & 2 & 11\\
{\it All articles} & {\it 30} & {\it 73} & {\it 56} & {\it 36} & {\it 22} & {\it 44}\\
\hline
\end{tabular}
{\small The count in each column is the number of subject categories
whose growth is within the given range.}
\caption{Histogram of change in fraction of older citations for broad areas. }
\label{tab:broad-change-histogram} 
\end{table}

\subsection{Change in growth rate}

Table~\ref{tab:broad-growth-two-step} presents the change in the
fraction of older citations over 1990-2001 and 2002-2013. Note that
the baseline for all growth percentages in the table is the fraction
of older citations in 1990. Using a common baseline allows us to
compare the growth percentages directly.

It shows that, overall, the increase in the second half was a little
more than double the increase in the first half. For all broad areas
that had a non-trivial increase in the fraction of older citations
over 1990-2013, the increase in the second half was substantially
larger than the increase in the first half. For 6 out of 9 areas, the
increase in the second half was at least double the increase in the
first half.

\begin{table}
\centering
\begin{tabular}{|l|r|r|}
\hline
Broad area & Change over 1990-2001 & Change over 2002-2013\\ \hline
Humanities, Literature \& Arts & 4\% & 14\%\\
Business, Economics \& Management & 19\% & 37\%\\
Social Sciences & 5\% & 26\%\\
Physics \& Mathematics & 11\% & 18\%\\
Life Sciences \& Earth Sciences & 11\% & 25\%\\
Engineering  & 2\% & 1\%\\
Chemical \& Material Sciences & -1\% & 3\%\\
Health \& Medical Sciences & 4\% & 26\%\\
Computer Science & 8\% & 31\%\\
{\it All articles} & {\it 9\%} & {\it 19\%}\\
\hline
\end{tabular}\\[0.25cm]
{\small Baseline for all growth percentages is the fraction of older citations in 1990. }
\caption{Change in the fraction of citations to older articles over 1990-2001 and 2002-2013.}
\label{tab:broad-growth-two-step}
\end{table}

\subsection{What about even older articles?}

\begin{figure}
\centering
\includegraphics{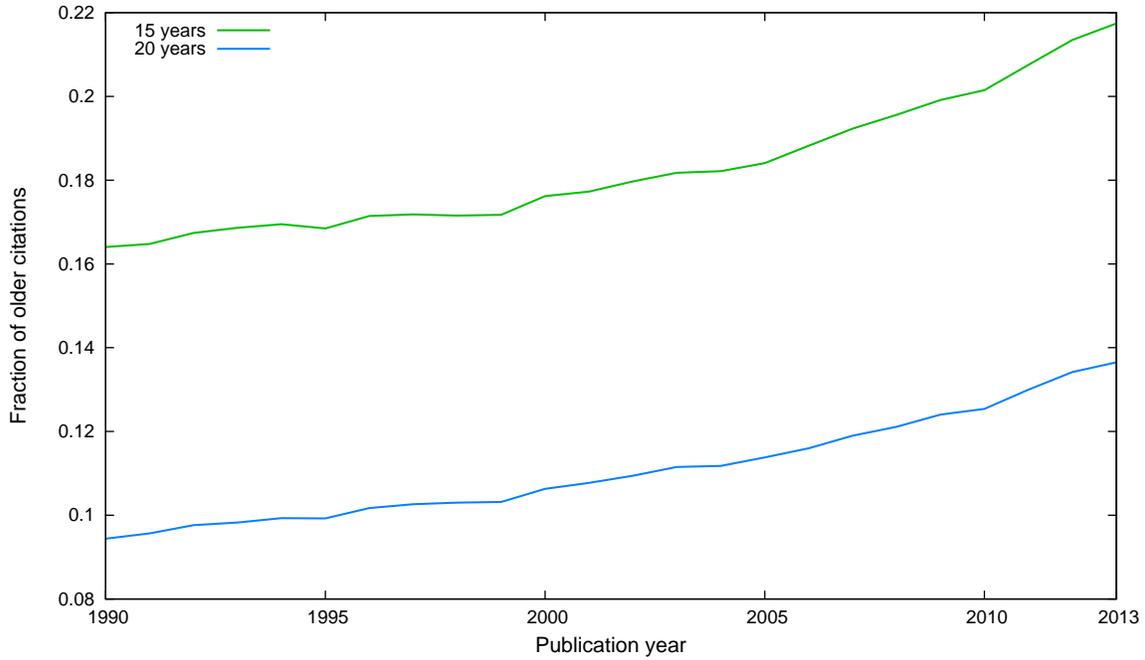}
\caption{Fraction of citations to even older articles across all papers. }
\label{fig:all-even-older} 
\end{figure}

Figure~\ref{fig:all-even-older} presents the change in the fraction of
older citations for two other definitions of ``older'' --- at least 15
years old and at least 20 years old. It shows that the fraction of
citations to even older articles grew continually over 1990-2013 and
that the growth for the second half (2002-2013) was significantly
larger than the growth for the first half (1990-2001).

\begin{figure}
\centering
\begin{tabular}{cc}
\includegraphics{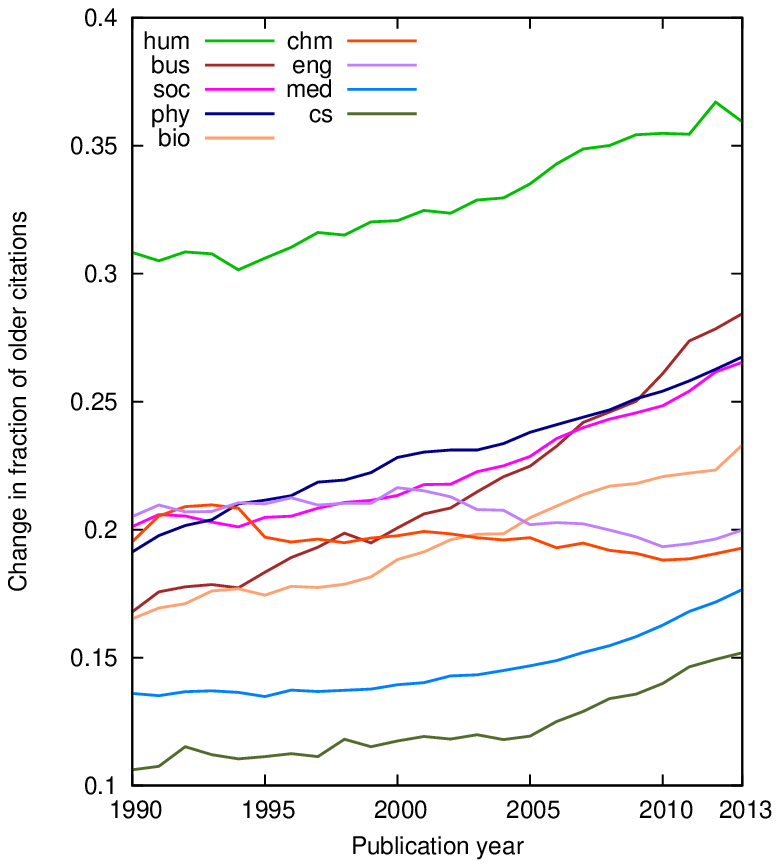} &
\includegraphics{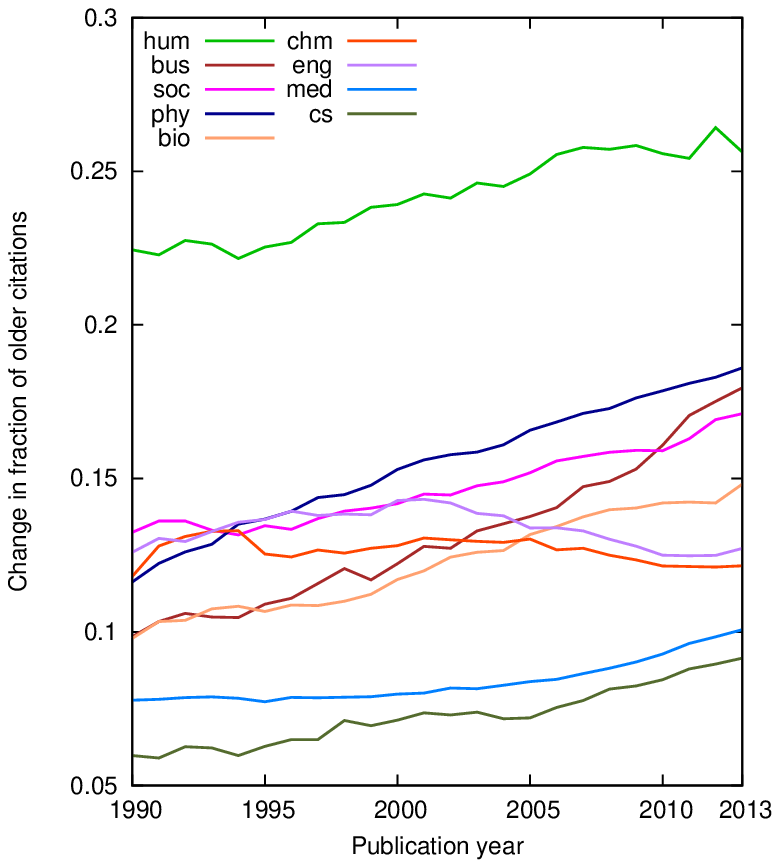} \\
{\small (a) $\geq 15$ years old}  & {\small (b) $\geq 20$ years old}\\
\end{tabular}\\[0.25cm]
{\small
{\bf bio:} Life Sciences \& Earth Sciences;
{\bf bus:} Business, Economics \& Management;
{\bf cs:} Computer Science;
{\bf chm:} Chemical \& Material Sciences;
{\bf eng:} Engineering;
{\bf hum:} Humanities, Literature \& Arts;
{\bf med:} Health \& Medical Sciences;
{\bf phy:} Physics \& Mathematics;
{\bf soc:} Social Sciences}\\
\caption{Fraction of citations to even older articles for broad areas of research. }
\label{fig:broad-even-older} 
\end{figure}

Figure~\ref{fig:broad-even-older} presents the evolution of the
fraction of citations to even older articles for all broad
areas. It shows that 7 out of 9 broad areas saw a
substantial increase in the fraction of citations to even older
articles over 1990-2013. Two broad areas, {\it Chemical \&
  Material Sciences} and {\it Engineering}, did not see a significant
change.

\section{Related Work}

The study of citation age in scholarly articles has had a long
history. Initial work in this area explored citation age as a way to
measure the rate of ``obsolescence'' in scholarly
literature~\cite{burton1960half,lawler1963psychology,line1970half,line1974progress,meadows1967citation,oliver1971effect}. One
of the goals of studying obsolescence was to provide guidance to
libraries regarding retention policies for older journal volumes.
Early metrics for citation age included
``half-life''~\cite{burton1960half} and ``Price's
Index''~\cite{price1970citation}. Line~\cite{line1970half} pointed out
the potential effect of the growth in the number of articles published
on citation age metrics -- if the number of articles grows rapidly,
one would expect the fraction of citations to recently published
articles to grow rapidly as well.

Exploring the notion of obsolescence from a usage perspective, rather
than a citation perspective, Sandison~\cite{sandison1974densities}
found that, after an initial period, the usage of older issues of
Physics journals at MIT didn't decrease with age.

In an early paper exploring the potential impact of online access on
scholarly communication, Odlyzko~\cite{odlyzko2002rapid} reported that
after an initial period, frequency of access to online articles from
several collections did not vary with age of articles. Based on this,
he speculated that easy online access to digitized collections, as
they become available, would lead to much wider usage of older
materials.

More recently, Evans~\cite{evans2008electronic} studied the impact of
online availability of journal articles on the age of citations. Based
on an analysis of citation indices from Thomson Reuters and a
database of online availability of journal articles from Information
Today Inc., he concluded that as more journal issues came online, the
articles referenced tended to be more recent. He speculated that the
shift from browsing print collections to searching online collections
facilitated avoidance of older literature. 

These results are in direct contrast with ours. We have found no
evidence that online availability leads to a reduction in citations to
older articles. To the contrary, we have found that, for most fields,
the growth in the fraction of older citations accelerated over the
period in which substantial numbers of articles became available
online and full-text searchable.

These results are also contradicted by two other studies that were
published around the same time as~\cite{evans2008electronic} and that
took different analysis approaches. Huntington
et~al~\cite{huntington2006article} studied article usage patterns
based on web access logs for OhioLink's journal collections. They
found that there were two stages in access history of scholarly
articles. The first stage spanned the first 8 to 9 years from
publication date. Usage often declined over this period, the decline
being sharpest in the first 2 to 3 years (a third over the first year
and by about 60\% by the third year). The next stage usually had a
relatively stable level of usage. Analyzing HTTP Referer headers for
journal article requests, they found that users arriving from search
services were far more likely to view older articles than users
arriving from a browse environment.  They speculated that this
difference occurred due to the relevance ranking approach used by web
search services.

Larivi{\`e}re et~al~\cite{lariviere2008long} studied the citations
from a large collection of articles published over 1900-2004. They
concluded that the useful life of scientific publications has been
increasing steadily since the 1970s. And that, in the aggregate over
all scholarship, and for natural sciences and engineering in
particular, the fraction of older citations has been steadily
increasing. Our results are in agreement with theirs.

\section{Conclusions}

There are three major conclusions from our study. First, the impact of
older articles, as measured by citations, has grown substantially over
1990-2013. Our analysis indicates that, in 2013, 36\% of citations
were to articles that are at least 10 years old and that this fraction
has grown 28\% since 1990. The fraction of older citations increased
over 1990-2013 for 7 out of 9 broad areas of research and 231 out of
261 subject categories.

Second, for most areas, the change over the second half (2002-2013)
was significantly larger than that over first half
(1990-2001). Overall, the increase in the second half was double the
increase in the first half.  Note that most archival digitization
efforts as well as the move to fulltext relevance-ranked search
occurred over the second half.

Third, the trend of a growing impact of older articles also holds for
articles that are at least 15 years old and those that are at least 20
years old. In 2013, 21\% of citations were to articles $\geq 15$ years
old with an increase of 30\% since 1990 and 13\% of citations were to
articles $\geq 20$ years old with an increase of 36\%.

In the introduction, we mentioned two broad trends that have the
potential to influence the fraction of older citations.  First,
finding and reading relevant older articles is now about as easy as
finding and reading recently published articles. This has made it
easier for researchers to cite the most relevant articles for their
work regardless of the age of the articles. Second, there has been a
dramatic growth in the number of articles published per-year. This has
significantly increased the number of recent articles that researchers
need to situate their work in relation to by citing.

Our results suggest that of the two trends, the ease of finding and
reading the most relevant articles, irrespective of their age, has had
the larger impact. For most fields, retrospective digitization as well
as inclusion in a broad-based search service with relevance ranking
occurred in the second half of the period of study. As mentioned
earlier, this is also the period that saw a larger growth in the
fraction of older citations.

Now that finding and reading relevant older articles is about as easy
as finding and reading recently published articles, significant
advances aren't getting lost on the shelves and are influencing work
worldwide for years after.

\bibliographystyle{plain}
\bibliography{paper}

\end{document}